\documentclass[%
 reprint,
]{revtex4-2}
\usepackage[top=3cm, bottom=3cm, left=3cm, right=3cm]{geometry}
\usepackage[utf8]{inputenc}
\usepackage{graphicx}
\usepackage{dcolumn}
\usepackage{bm}
\usepackage{comment}
\usepackage{hyperref}
\usepackage{amsmath}
\usepackage{amssymb}
\usepackage{xcolor}
\usepackage{listings}
\usepackage{cleveref}

\usepackage{subfig}
\usepackage{float}
\usepackage{array,multirow}
\usepackage[nolist]{acronym}
\usepackage{float}
\makeatletter
\edef\alg@ftype{\the\value{float@type}} 
\makeatother

\usepackage{algorithm}
\usepackage{algpseudocode}

\makeatletter
\expandafter\let\csname ftype@algorithm\endcsname\alg@ftype 
\makeatother
\usepackage{placeins}

\graphicspath{{./figures/}}

\def \beq {\begin{eqnarray}}
\def \eeq {\end{eqnarray}}

\def \bra {{\langle}}
\def \ket {{\rangle}}

\begin{document}
\title{Selected Stabilizer Subspaces for Compact Many-Body Wave Functions}
\author{Maria-Andreea Filip}\email{maf63@cam.ac.uk} 

\affiliation{Yusuf Hamied Department of Chemistry, University of Cambridge,
Cambridge CB2 1EW, UK }

\author{George H. Booth}
\affiliation{Department of Physics and Thomas Young Centre, King’s College London, Strand, London, WC2R 2LS, UK}
\date{\today}
\begin{abstract}
    Selected configuration interaction (SCI) obtains accurate many-body solutions by variationally diagonalising the Hamiltonian in a sparse, iteratively selected set of configurations, but can lose compactness when the target state has broad configurational support. We introduce selected stabilizer interaction (SSI), a fully classical, SCI-inspired method that instead iteratively constructs a nonorthogonal subspace spanned by a systematically enlargeable set of selected stabilizer states. A single stabilizer, generated by Clifford operations, can encode a structured superposition of (possibly exponentially) many configurations, shifting the relevant compression from configurational to stabilizer sparsity. We benchmark ground-state energies, fidelities, superposition degrees and required expansion lengths against exact diagonalization and equivalent SCI schemes for the transverse-field Ising, XXZ Heisenberg, and Hubbard models, finding particularly strong compression from the stabilizer representation in the spin models, and increasing advantage with system size for these systems. The results establish a proof of principle for heuristic selected stabilizer diagonalization, while also identifying particular challenges in symmetry handling and effective exploration of the stabilizer space.
\end{abstract}
\maketitle

\newcounter{dummy} 
\newcommand{\diff}{\mathop{}\!\mathrm{d}} 
\newcommand{\Supp}{\mathop{}\!\mathrm{Supp}}
\newcommand{\deriv}[3][]{
\frac{\diff^{#1}#2}{\diff #3^{#1}}}
\newcommand{\pderiv}[3][]{%
\frac{\partial^{#1}#2}
{\partial #3^{#1}}}
\newcommand{\vect}[1]{\mathbf{#1}} 
\newcommand{\mat}[1]{\mathbf{#1}}
\newcommand{\blankpage}{\newpage\hbox{}\thispagestyle{empty}\newpage}
\newcommand{\hairsp}{\hspace{1pt}}
\newcommand{\ie}{\textit{i.\hairsp{}e.}\xspace}
\newcommand{\Ie}{\textit{I.\hairsp{}e.}\xspace}
\newcommand{\eg}{\textit{e.\hairsp{}g.}\xspace}
\newcommand{\Eg}{\textit{E.\hairsp{}g.}\xspace}
\newcommand{\vs}{\textit{vs.}\xspace}
\newcommand{\abs}[1]{\left|#1\right|}
\newcommand{\braket}[2]{\left\langle#1|#2\right\rangle}
\newcommand{\order}[1]{\mathcal{O}\left(#1\right)}
\newcommand{\normord}[1]{n[#1]}
\newcommand{\normordf}[1]{\{#1\}}
\newcommand{\normordop}[1]{\hat{#1}_{\textrm{N}}}
\newcommand{\commu}[2]{\left[#1,#2\right]}

\newcommand{\crea}[1]{a_{#1}^\dagger}
\newcommand{\annih}[1]{a_{#1}}
\newcommand{\bcrea}[1]{b_{#1}^\dagger}
\newcommand{\bannih}[1]{b_{#1}}
\newcommand{\rpamom}[1]{\mat{\eta}^{(#1)}}
	
\begin{acronym}
	\acro{RPA}{Random Phase Approximation}
	\acro{ERPA}{Extended Random Phase Approximation}
	\acro{QBA}{Quasi-Boson Approximation}
	\acro{RPAX}{Random Phase Approximation with Exchange}
	\acro{dRPA}{direct Random Phase Approximation}
	\acro{FCI}{Full Configuration Interaction}
	\acro{RDM}{Reduced Density Matrix}
	\acro{RDMs}{Reduced Density Matrices}
	\acro{HF}{Hartree--Fock}
	\acro{CAS}{Complete Active Space}
	\acro{CI}{Configuration Interaction}
	\acro{NO}{Natural Orbital}
	\acro{DMET}{Density Matrix Embedding Theory}
	\acro{EwDMET}{Energy-weighted Density Matrix Embedding Theory}
	\acro{EoM}{Equation of Motion}
	\acro{SVD}{Singular Value Decomposition}
	\acro{DFT}{Density Functional Theory}
	\acro{TDM}{Transition Density Matrix}
	\acro{RI}{Resolution of the Identity}
	\acro{EDMET}{Extended Density Matrix Embedding Theory}
	\acro{DMFT}{Dynamical Mean Field Theory}
	\acro{TDDFT}{Time-Dependent Density Functional Theory}
	\acro{DFT}{Density Functional Theory}
	\acro{PPP}{Pariser--Parr--Pople}
	\acro{MRSDCI}{multireference singles-doubles configuration-interaction}
	\acro{CO}{charge ordered}
	\acro{AFM}{anti-ferromagnetic}
	\acro{EHM}{extended Hubbard model}
	\acro{BOW}{bond order wave}
	\acro{DMRG}{density matrix renormalisation group}
	\acro{DDMRG}{dynamical DMRG}
	\acro{SDW}{spin density wave}
	\acro{CDW}{charge density wave}
	\acro{PSD}{positive semi-definite}
	\acro{NI}{numerical integration}
	\acro{cRPA}{constrained RPA}
	\acro{CAS}{complete activate space}
	\acro{CASSCF}{complete active space self-consistent field}
	\acro{TDA}{Tamm--Dancoff approximation}
\end{acronym}

\section{Introduction}

The exponential growth of Hilbert space remains a central obstacle in quantum many-body physics. Primarily in quantum chemistry, selected configuration-interaction (SCI) methods have emerged as a leading route beyond this obstacle for high-accuracy approximations in many relatively small systems~\cite{Huron1973,Tubman2016,Holmes2016,Holmes2016b,10.1021/acs.jctc.8b00536,10.1063/1.4955109}. By retaining a carefully chosen expansion of only a small fraction of the full determinant space, they can approach full CI accuracy for systems far beyond brute-force diagonalization and are increasingly used to generate benchmark data for less controlled approximations for small molecules~\cite{PhysRevX.10.011041,10.1021/acs.jpclett.0c02621}, and have even been applied to some lattice and impurity problems~\cite{PhysRevB.100.125165,zhang2025randomdeterminantsgroundstate}.

The premise of SCI is simple. If a target eigenstate is sufficiently compressible to a sparse representation in the configurational basis, the Hamiltonian can be projected into this low-rank subspace spanned by its most important Slater determinants and diagonalized. The projected ground-state energy is variational, and the approximation can be systematically improved by enlarging the selected space. Success rests on efficient classical heuristics to find the relevant configurations without enumerating the full Hilbert space, and efficient evaluation of the required Hamiltonian matrix elements. For many systems, an appropriate Hartree–Fock, natural-orbital, or otherwise optimized one-particle representation renders the target state sufficiently sparse that large amounts of configurational “deadwood” can be discarded with little loss of accuracy, and the method performs well~\cite{Ivanic2001}. Methods such as CIPSI, heat-bath CI and other variants differ mainly in how they initialize, enlarge, and sometimes prune the variational space, with candidate configurations drawn from the first-order interacting space of the current approximation and ranked using estimates of their amplitudes, Hamiltonian couplings, or perturbative energy contributions. Related determinant-space compression under assumptions of a sparse many-body state also appear in the stochastic approach of full configuration-interaction quantum Monte Carlo~\cite{10.1063/1.3193710}, while more recently, sample-based quantum diagonalization and related workflows have used configurations selected from the Born distribution of a quantum-prepared state to define a subspace for subsequent classical diagonalization~\cite{Kanno2026,Robledo-Moreno2025,Motta2023,Liepuoniute2025,Shajan2025,Shajan2026,Reinholdt2025,Yu2025,Piccinelli2026,Barison2025}.

Despite their different selection mechanisms, these approaches all share a common bottleneck: the target state must remain compact in terms of the rank of the selected configurational expansion. Multireference and strongly correlated systems where large-scale superposition is required, or where the number of important configurations grows extensively or exponentially, can erode this advantage, with the costs of discovering, storing, and diagonalizing the selected space eventually proving too much of a bottleneck. This limitation motivates this work, where we retain the selected-subspace principle, but change the class of states from which that subspace is constructed to allow large-scale superposition of configurations while remaining classically tractable.

We do this by replacing the selection of individual many-body configurations by stabilizer states. A stabilizer state is a Clifford transform of a computational-basis state and admits a compact efficient classical representation~\cite{Aaronson2004}. Stabilizer states underpin efficient Clifford-circuit simulation and the stabilizer formalism of quantum error correction~\cite{Gottesman1997,Gottesman1998,Bravyi2016,Bennink2017,Bravyi2019}, but single stabilizer states can also describe highly structured many-body entanglement, and may have support on exponentially many configurations, in principle enabling large-scale superposition within compact subspace expansions. Nevertheless, matrix elements of low-body operators between stabilizer states remain efficiently classically computable without expanding them explicitly in their constituent configurations.

This means that the relevant notion of sparsity required for this approach is then stabilizer rather than configurational compressibility, which are not necessarily the same. The stabilizer rank of a state is the minimum number of stabilizer states required for an exact decomposition, while approximate stabilizer rank incorporates a specified error tolerance, and is therefore the relevant quantity to probe the efficiency of such an approach~\cite{PhysRevX.6.021043,deSilva2024}. These quantities are closely connected to the non-stabilizerness or a resource known as “magic” for a state~\cite{Veitch_2014,Bravyi2019,Oliviero2022,Sarkis2025}. Relevant many-body states with broad configurational support may admit compact stabilizer expansions~\cite{PhysRevLett.86.910,PhysRevA.69.062311,KITAEV20032}. Moreover, every computational-basis configuration is itself a stabilizer state. Therefore, with mild assumptions, the optimal stabilizer ansatz of a given rank can do no worse than the optimal equivalent length configurational ansatz. Whether this formal increase in expressiveness can be converted into a practical reduction in subspace size is the key algorithmic question we investigate in this work, in the development of a `selected stabilizer interaction' (SSI). 

Recent related work has also explored stabilizer and more general low-magic representations of quantum many-body ground states~\cite{seibert2026correlationmagicelectronicstructure}. Single stabilizer state approximations have been investigated for condensed-matter and electronic-structure Hamiltonians~\cite{Sun2025,5qr5-7jkz,PhysRevA.111.062413,Mao2026}, while complementary work has developed algorithms for perturbations from stabilizer Hamiltonians~\cite{PhysRevA.110.062427}. Stabilizer structure has also been combined with tensor-network representations~\cite{PhysRevLett.133.230601,PhysRevLett.133.190402}. Most closely related to the present work, Anand and Brown~\cite{jkcp-6km5} introduced an excitation-guided search for low-energy single-stabilizer approximations to molecular ground states, together with a restricted, one-parameter superposition of two Pauli-related stabilizers. This contrasts with our more general systematic expansion of non-orthogonal stabilizers of this work. 

At the operator level, complementary Pauli-propagation approaches also exploit sparsity in low-magic representations of Hamiltonians under successive transformations~\cite{Rall2019}, which has also been recently applied to ground-state energy estimation for strongly correlated Heisenberg and Hubbard models~\cite{Shrikhande2026}. Finally, we also note that changing the underlying basis state definitions from which a selected CI-like expansions can be iteratively searched was also investigated in Ref.~\cite{10.1021/acs.jctc.0c00141}, where tensor products of superpositions of states over clusters of orbitals was considered, again building superposition and correlation into the definition of the underlying basis from which states were searched and iteratively selected. Also related are non-orthogonal CI expansions, where the overcompleteness and redundancy of the Slater determinant space is exploited by combining each determinant with a basis rotation or fermionic Gaussian unitary operation. This additional flexibility in the basis can be used to develop heuristic algorithms for highly compact classical expansions with high accuracy~\cite{Giuliani2026}. In this work, we are considering the use of Clifford rather than Fermionic Gaussian unitaries, as classically tractable circuits to add flexibility into the basis for these expansions.


Stabilizer states form an enormous, overcomplete, and nonorthogonal set, requiring the development of new heuristic approaches to search this candidate space effectively, control the nonorthogonality and realize the representational benefit. We therefore develop a classical heuristic for iteratively searching, selecting, and refining stabilizer subspaces for subsequent variational many-body diagonalization. Inspired by the selection principles of SCI, the method identifies stabilizer states that improve the current approximation and repeatedly solves the Hamiltonian within their span. We apply the resulting framework to condensed matter applications where SCI would often struggle, the transverse-field Ising, Heisenberg, and Hubbard models, for which its accuracy can be rigorously benchmarked on small lattices. We assess both the quality of the resulting variational ground-state solutions and the compactness of their stabilizer expansions relative to configuration-selected approaches. Our results identify regimes in which stabilizer superpositions can provide a significantly more compact description of the target state, particularly in the high external field regime of the transverse field Ising model where the solution tends to a single stabilizer state, while exposing the challenges that remain to realize an advantage in fermionic systems.

\section{Review of Stabilizer states} \label{sec:theor}

We begin by defining stabilizer states and giving a pedagogical review of some of their properties relevant for this work, with more details found in Refs. \cite{Gottesman1997,Garcia2014}.
A \textit{stabilizer} for a state $|\Psi\ket$ is any operator $\hat U$ such that
\begin{equation}
    \hat U |\Psi\ket = |\Psi\ket.
\end{equation}
As trivial examples, the identity $I$ stabilizes every quantum state, while $-I$ does not stabilize any state. The set of all operators that stabilize a given state forms the \textit{stabilizer group} of that state. Restricting this group to Pauli operators, we denote
\begin{equation}
    \mathcal{S}(\Psi) = \{\hat P \in \mathcal{P}_n \ | \ \hat P|\Psi\ket = |\Psi\ket\},
\end{equation}
where $\mathcal{P}_n$ is the full group of phased $n$-qubit Pauli operators and $|\Psi\ket$ is an $n$-qubit quantum state. A \textit{stabilizer state}, $|\sigma\ket$ is any $n$-qubit quantum state that is stabilized by exactly $2^n$ mutually commuting Pauli operators, i.e. $|\mathcal{S}(\sigma)| = 2^n$. Despite being exponentially large, this group admits linear-size generating sets: any $n$ linearly independent non-identity Pauli strings from $\mathcal{S}(\sigma)$ will generate the whole group. Therefore, stabilizer states may be efficiently encoded classically by specifying this generating set, knowledge of which is sufficient to uniquely identify them. Despite this efficient classical description, stabilizer states may have exponentially large support on the computational basis states or be highly entangled. Example stabilizer states to demonstrate these properties are the uniform product state consisting of the exponential superposition over all $n$-qubit computational basis states, 
    \begin{equation}
        |+\rangle^{\otimes n} =\frac{1}{\sqrt{2^n}}\sum_{j=0}^{2^n -1} |j\ket ,
    \end{equation}
    stabilized by the group generated by $\{X_j | j \in \{0, 1,...,n-1\}\}$, and the maximally entangled $n$-qubit GHZ state,
    \begin{equation}
        \frac{1}{\sqrt 2}(|0\ket^{\otimes n} + |1\ket^{\otimes n}),
    \end{equation}
    stabilized by the group generated by $\{X^{\otimes n}\} \cup \{Z_iZ_{i+1} | i \in \{0, 1,..., n-2\}\}$.
    A useful property of stabilizers is that given two orthogonal stabilizer states, where one is generated by the action of a Pauli string, $\hat P$, on the other, i.e. $\langle \sigma |\hat  P | \sigma \rangle=0$, then the state
\begin{equation}
    |\sigma_\pm\ket = \frac{1}{\sqrt 2}( I \pm \hat P) |\sigma\ket
    \label{eq:stab_nn}
\end{equation}
is also a stabilizer state. 

There are two common representations for the generating set of a stabilizer group, which we will exemplify here using the 4-qubit GHZ state. The first uses the Pauli operators themselves as elements, so the generator set is represented by an $n \times n$ matrix, with each row corresponding to an element of the stabilizer generators. For the GHZ state, this is
\begin{equation}
    \mathbf{M}_\mathcal{G} = \begin{pmatrix}
        X & X & X & X\\
        Z & Z & I & I\\
        I & Z & Z & I\\
        I & I & Z & Z
    \end{pmatrix}.
\end{equation}

Another approach is to use the symplectic representation of the Pauli operators, where each is represented as two binary length-$n$ strings over the vector space $\mathbb{F}_2^{2n}$ (ignoring global phases for the moment). In this case, the action on each qubit is represented by a two-bit number, with the first bit corresponding to $X$ action and the second to $Z$ action, i.e. $I = 00$, $Z = 01$, $X = 10$, $Y = 11$ (up to a phase).
By writing each Pauli string as a length $2n$ binary vector $(\mathbf{x}|\mathbf{z}) \in \mathbb{F}_2^{2n}$, Pauli string multiplication can be computed as binary addition of the bit strings, while the binary symplectic inner product of these bit strings can efficiently compute the commutator of two Pauli operators.
The symplectic form of the stabilizer generator matrix is therefore an $n \times 2n$ binary matrix, with the first $n$ columns corresponding to the `$X$ bits' and the second to the `$Z$ bits'. A requirement of a valid generator set of a stabilizer state is that they mutually commute, hence the rows are pairwise orthogonal under the binary symplectic inner product. Furthermore, the linear independence of the generators enforces that the rank of the generator matrix under $\mathbb{F}_2$ arithmetic is equal to $n$. For the GHZ state, this is
\begin{equation}
    \mathbf{M}_\mathcal{G} = \left(\begin{array}{cccc|cccc}
        1 & 1 & 1 & 1 & 0 & 0 & 0 & 0\\
        0 & 0 & 0 & 0 & 1 & 1 & 0 & 0\\
        0 & 0 & 0 & 0 & 0 & 1 & 1 & 0\\
        0 & 0 & 0 & 0 & 0 & 0 & 1 & 1
    \end{array}\right).
\end{equation}
In both representations, the phases associated with each Pauli operator must also be stored and considered appropriately in all operations. 
Once generators are specified, other elements of the stabilizer group may be generated, either by row multiplication of Pauli operators, or by binary addition (\textsc{xor}) in the symplectic representation. 

There are a vast number ($\mathcal{O}[2^{n^2}]$) of 
equivalent ordered generating sets for the same stabilizer group,\cite{Aaronson2004} so we choose a unique canonical representation which we will label $\mathcal{G}(S)$.
The canonicalization we use minimises both the number of non-diagonal ($X$ or $Y$ containing) generators and the number of $Z$ operators in the diagonal (only $Z$ and $I$ containing) generators. 
Canonicalisation of the generator matrix corresponds to a Gaussian elimination procedure, which preserves the rank and pairwise commutativity of the generators, and provides the unique representation, with the detailed algorithm given in Ref. \cite{Garcia2014}. 
The total number of $n$-qubit stabilizer states is given by
\begin{equation}
    \mathcal{N}(n) = 2^n\prod_{k=0}^{n-1}(2^{n-k} +1) \sim \mathcal{O}[2^{n^2}].
\end{equation}
This exposes the redundancy in the dimension of the stabilizer basis over the configurational space which scales super-exponentially in system size.

Stabilizer states may also be represented by the action of a \textit{stabilizer circuit} $\mathcal{C}(\sigma)$ on the all-zero state, such that
\begin{equation}
    \mathcal{C}(\sigma)|0\ket^{\otimes n} = |\sigma\ket.
\end{equation}
This circuit is composed entirely of Clifford operations ({\sc Cnot}, Hadamard and Phase gates, which also generate all Pauli gates), which can be simulated classically via Gottesman-Knill methods.\cite{Gottesman1997,Gottesman1998b} Various canonical circuit constructions have been proposed;\cite{Aaronson2004, VanDenNes2010, Garcia2014}in this work we follow the construction of Garc\'ia \textit{et al.}, which uses a 5-block sequence $H - CX - CZ - P - H$~\cite{Garcia2014}.
Knowledge of the canonical generator matrix and/or stabilizer circuit corresponding to different stabilizer states allows efficient computation of the overlap between them~\cite{Garcia2014}. While computing the phaseless overlap is relatively simple, including the phase can prove challenging, and is often neglected in many descriptions in the literature. We therefore give our algorithm for the phased overlap between stabilizers, adapted from Ref. \cite{Garcia2014}, in Algorithm \ref{alg:overlap} in \Cref{app:code}.

Once the phased overlap can be found, this naturally leads to the evaluation of Hamiltonian matrix elements between stabilizers. For a Hamiltonian expressed as a linear combination of Pauli strings, $\hat H=\sum_k h_k \hat P_k$, its matrix elements within a selected stabilizer subspace reduce to phase-sensitive stabilizer overlaps,
\begin{equation}
H_{\alpha\beta}=\sum_k h_k
\langle\sigma_\alpha|\hat P_k|\sigma_\beta\rangle,
\end{equation}
since $\hat P_k|\sigma_\beta\rangle$ is itself a stabilizer state. Consequently, both the Hamiltonian matrix and the overlap matrix $S_{\alpha\beta}=\langle\sigma_\alpha|\sigma_\beta\rangle$ can be evaluated classically in polynomial time. The overcompleteness of the stabilizer set provides substantial freedom to construct compact approximations (akin to how non-orthogonal determinant expansions can result in compact representations~\cite{Giuliani2026}), but also necessitates careful treatment of linearly dependent, or poorly conditioned selected states~\cite{deSilva2024}.

\section{Selected Stabilizer Interaction} \label{sec:ssi}
\subsection{Motivation} \label{sec:motivation}
The main aim of this paper is to ascertain whether (i) stabilizer states can provide a more compact and classically tractable representation of physical wavefunctions than conventional configurational expansions and (ii) whether increasingly large expansions may be found effectively through a systematically improvable SCI-like algorithm. We note that any two computational basis states which have coefficients of the same magnitude in the target wavefunction may be grouped together in a single stabilizer state, according to \Cref{eq:stab_nn}. However, in some cases, $2^m$ configurations may be grouped together where $m$ is an integer between 0 and $n$, provided they form an affine subset of $\mathbb F_2^n$, and have structured coefficients with equal magnitude~\cite{Dehaene2003,deSilva2024}. We refer to $m$ as the affine subspace dimension or `superposition degree' with $2^m$ as the size of the support for a stabilizer state over the configurational basis, i.e.
\begin{equation}
    m(|\sigma\rangle) = \log_2 | \{ x \in \{0,1\}^n : \langle x | \sigma \rangle \neq 0 \}| . \label{eq:supdeg}
\end{equation}
This value is also equal to the rank of the $X$-block of the stabilizer tableau. We exemplify stabilizer compression with a Transverse Field Ising (TFI) model Hamiltonian,
\begin{equation}
    \hat H_\text{TFI} = J \sum_{i =0}^{n-2} Z_iZ_{i+1} + h \sum_{i=0}^{n-1} X_i .
    \label{eq:tfi}
\end{equation}
By global spin-flip and spatial reflection symmetries of the model (with generators $X_0X_1X_2$ and $X_0 Z_1 X_2$), the $+1$ and reflection-even symmetry many-body states for $n=3$ must have the form
\begin{equation}
\begin{split}
        |\Psi\ket = &a (|000\ket + |111\ket) + b (|010\ket + |101\ket)\\
        + &c(|001\ket + |100\ket + |110\ket + |011\ket).
\end{split}
\end{equation}
This may be rewritten as a sum of three stabilizer states:
\begin{equation}
   |\Psi\ket = \sqrt{2}a|S_1\ket + \sqrt{2}b|S_2\ket + 2c|S_3\ket, 
\end{equation}
so the exact stabilizer rank for these states is at most three, compared to a computational basis dimension of eight. The aim of SSI is to recover such expansions in a robust black-box fashion, and we hope that approximate groupings that give effective compressions of the expansion rank relative to configurational space may be also found in broader settings, without relying on explicit symmetry-derived compressions.

\subsection{Methodology}\label{sec:ssi_algo}
\begin{figure*}
    \centering
    \includegraphics[width=\linewidth]{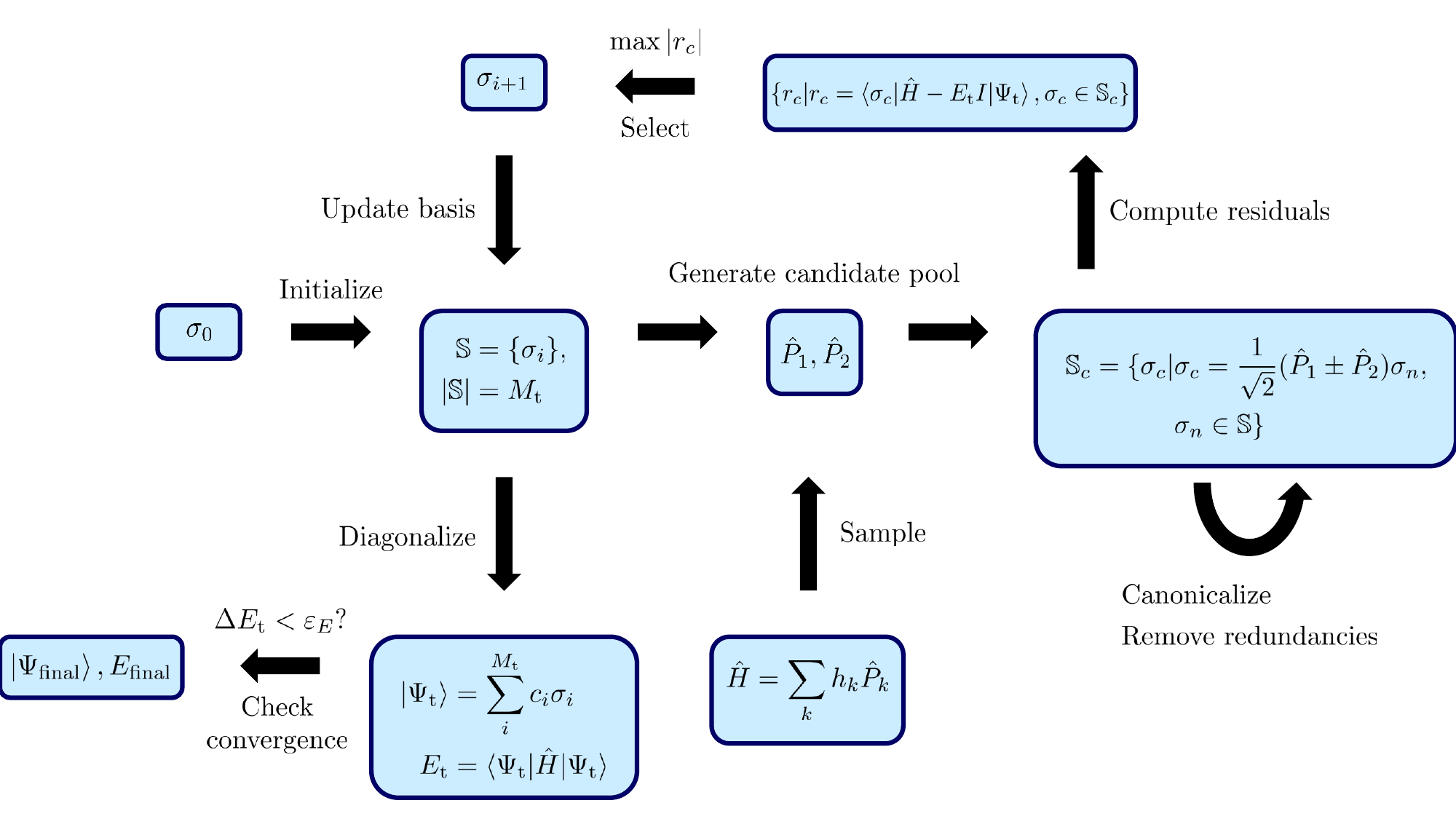}
    \caption{Schematic representation of the SSI algorithm. The stabilizer basis is initialized from a single `seed' state $\sigma_0$. Candidates are then generated as described in \Cref{sec:cg} and ranked according to a residual as described in \Cref{sec:sc}. The top candidate is added to the stabilizer basis and a new trial wavefunction is obtained by diagonalization (see \Cref{sec:d}). The process is repeated until the energy change is below some convergence threshold.}
    \label{fig:algo-flowchart}
\end{figure*}
We propose a simple SSI algorithm, shown schematically in \Cref{fig:algo-flowchart}, and broadly motivated by SCI methodologies, with a few important features necessary to take advantage of the stabilizer space structure.

\subsubsection{Candidate generation}
\label{sec:cg}
As the space of stabilizer states grows super-exponentially with the number of qubits, an exhaustive search of all possible stabilizers in order to select the optimal truncated expansion is an even more intractable task than it is in conventional SCI. Even the number of distinct stabilizers that have non-zero overlap with $\hat H|\sigma\rangle$, where $|\sigma \rangle$ is a single stabilizer state, will generally form a super-exponentially scaling overcomplete set due to the high degree of overlap between the states. We therefore stochastically generate a small candidate pool of stabilizers at each iteration from a current trial state, \mbox{$|\Psi_\text{t}\rangle=\sum_i c_i |\sigma_i \rangle$}. From this pool, we evaluate a selection criterion to determine the inclusion of states in the expansion. Ideally, we would like to select new stabilizers with large Hamiltonian coupling to the current trial state, $|\bra \sigma_n|\hat H|\Psi_\text{t}\ket|$. As a heuristic approach, we implement the following procedure:
\begin{enumerate}
    \item Select $N_\text{att}$ \textit{pairs} of signed Pauli terms $\{s_j \hat P_j, s_k\hat P_k\}$ in the Hamiltonian, with probability $p_k \propto |h_k|$, where $\hat H = \sum_k h_k \hat P_k$, and $s_k=\textrm{sgn}(h_k)=\pm1$ is a binary sign variable assuming real-valued Hamiltonians.
    \item Iterate through the stabilizer states $|\sigma_i\ket$ present in the expansion of $|\Psi_\text{t}\ket$. For larger systems, it would be possible to stochastically select a number of stabilizer states $\propto |c_i|^2$, in order to avoid the complete enumeration of the current basis of stabilizers in the trial state.
    \item For every selected stabilizer, $|\sigma_i\rangle$ from $|\Psi_\text{t}\rangle$, and sampled pair of Pauli terms from $\hat{H}$, we construct the new candidate stabilizer state
    \begin{equation}
        |\sigma_c\ket = \frac{1}{\sqrt{2}}\left.\left.\left(s_j\hat P_j + s_k\hat P_k\right)\right|\sigma_i\right\rangle.
        \label{eq:cand_stab}
    \end{equation}    
\end{enumerate}
This may be rewritten by defining
\begin{equation}
    A=s_j\hat P_j,\ \
    B=s_k\hat P_k,\ \
    Q=AB=s_js_k\hat P_j\hat P_k,
\end{equation}
such that
\begin{equation}
    |\sigma_c\rangle
    =
    A\frac{I+Q}{\sqrt{2}}|\sigma_i\rangle.
\end{equation}
The vector $|\sigma_c\rangle$ either vanishes or is
proportional to a stabilizer state. There are three cases:
\begin{enumerate}
\item If $\{A,B\}=0$, then $Q^\dagger=-Q$, and consequently
\begin{equation}
    U=\frac{I+Q}{\sqrt{2}}
\end{equation}
is a unitary Clifford operator by \Cref{eq:stab_nn}, and
$|\sigma_c\rangle=AU|\sigma_i\rangle$ is an already
normalized stabilizer state.
    \item If $[A,B]=0$ and $Q\in\mathcal S(\sigma_i)$, then
\begin{equation}
    |\sigma_c\rangle
    =\sqrt{2} A|\sigma_i\rangle.
\end{equation}
After normalization, the candidate is therefore $A|\sigma_i\rangle$.
If instead $-Q\in\mathcal S(\sigma_i)$, then
$|\sigma_c\rangle=0$ and the candidate is discarded.
    \item If $[A,B]=0$ and neither $Q$ nor $-Q$ belongs to
$\mathcal S(\sigma_i)$, then $(I+Q)/2$ is the projector onto the
$+1$ eigenspace of the Hermitian Pauli operator $Q$. In this case
$\langle\sigma_i|Q|\sigma_i\rangle=0$, so
$|\sigma_c\rangle$ is normalized and is a stabilizer state
obtained by the standard Pauli-measurement update.
\end{enumerate}
Every nonzero candidate is normalized and canonicalized, with duplicates removed before subsequent ranking.
Generating new candidates according to \Cref{eq:cand_stab} allows candidate stabilizer states to change superposition degree, $m$, compared to their parent stabilizer, which is the main motivation for use of stabilizer expansions. They therefore encode a different number of constituent configurations in their superposition, and allow this property to evolve during the course of the simulation. This would not be possible from the action of a single Pauli string which would conserve $m$, and is essential to drive exploration of the the desired compression derived from this basis.

We note that by restricting the selection to a single Pauli operator per step and starting from a computational basis state, this algorithm reduces to a simple SCI method. 
Therefore, to answer our key question as to the compactness of these heuristic selection schemes as we move from computational basis states to stabilizer states, in the SCI results of \Cref{sec:results} we can compare to this matched SCI method, where we use equivalent proposals, residual ranking heuristics and one-state-at-a-time updates for fairness of comparison.

\subsubsection{Selection criterion}
\label{sec:sc}
Once candidates have been generated, they are ranked according to each candidate's projection onto the Ritz residual criterion
\begin{align}
    \begin{split}
        r_n &= |\bra \sigma_n|\hat H - E_\text{t}|\Psi_\text{t}\ket|\\
            &= \left|\sum_i^{M_\text{t}} c_i (H_{ni} - E_\text{t}S_{ni})\right|,
    \end{split}
    \label{eq:residual}
\end{align}
where $E_\text{t}$ is the energy corresponding to the current trial state, and $S_{ni}=\langle \sigma_n | \sigma_i \rangle$. This criterion only considers stabilizer states not currently explicitly included in the variationally optimized trial state, and is small if $|\sigma_n\ket$ has significant overlap with the current state, and increases with the strength of the Hamiltonian coupling. Alternatives which include energy-denominators to estimate energy lowering have been used in other SCI approach and will be considered in future work. 

Ranking all candidates requires computation of Hamiltonian and overlap matrix elements between all stabilizers in the trial state and the candidate pool, which is the most computationally intensive part of this algorithm, but which can also be controlled in each iteration by limiting the candidate pool size. Once these residuals have been computed, one could select the top $N$ contributors or implement a threshold and select all $r_i > \epsilon$. However, for simplicity in the results presented here, we only include the top $N=1$ stabilizer at each iteration to allow for a more optimal (yet perhaps slower) growth of the expansion.

\subsubsection{Energy evaluation}
\label{sec:d}
We find the new energy estimate and wavefunction by solving the generalised eigenvalue problem in the space of the stabilizers in the original trial and selected candidates,
\begin{equation}
    \mathbf{Hc} = E \mathbf{Sc},
\end{equation}
where $H_{ij} = \bra \sigma_i|\hat H |\sigma_j\ket$ and $S_{ij} = \bra \sigma_i|\sigma_j\ket$. Convergence is considered achieved once the energy change in successive iterations falls below a predefined tolerance, at which point we have a final energy expectation and wave function as a linear combination of stabilizer states. Singular and near-singular directions in the overlap matrix $\mathbf{S}$ are handled by setting a linear dependency threshold on the eigenvalues of $\mathbf{S}$. Any step that generates a singular overlap matrix within this threshold is discarded.

Most operations on the generator matrix of a stabilizer state computationally scale as $\mathcal{O}(n^3)$, as they require re-canonicalisation, which is a Gaussian elimination process. Conjugation by a Pauli string, as required to apply Hamiltonian terms, scales as $\mathcal{O}(n^2)$. Candidate generation therefore scales as $\mathcal{O}(N_\text{att}M_\text{t}n^2)$, where $M_\text{t}$ is the size of the stabilizer expansion.  Computing the residuals scales as $\mathcal{O}(N_\text{att}N_hM_\text{t}^2n^3)$, where $N_h$ is the number of terms in the Hamiltonian. This cost can be reduced by caching stabilizer overlaps computed on previous steps. To balance memory usage and speed, we cache $S_{ij}$ where $\sigma_i$ is in the current trial wavefunction and $\sigma_j$ is in $\hat H |\Psi_\text{t}\ket$. We do not store overlaps involving rejected candidates or states connected to them by the Hamiltonian. Finally the diagonalization step scales as $\mathcal{O}(M_\text{t}^3)$. The bottleneck therefore shifts from the residual computation to the diagonalization step as the number of stabilizers in the expansion increases, though this could be reduced in the future if necessary with direct and iterative techniques.

\section{Numerical Results} \label{sec:results}

We benchmark our method for model spin Hamiltonians (the Transverse-field Ising (TFI) model, the Heisenberg XXZ model), as well as the fermionic Hubbard model. These lattice models are traditionally considered to be outside the scope of SCI methods due to their significant delocalization of the state across the configurational space at relevant Hamiltonian parameter regimes, and we compare to SCI explicitly to consider the compactness of the resulting expansions.

\subsection{Transverse field Ising model}

\begin{figure*}
    \centering
    \includegraphics[width=\linewidth]{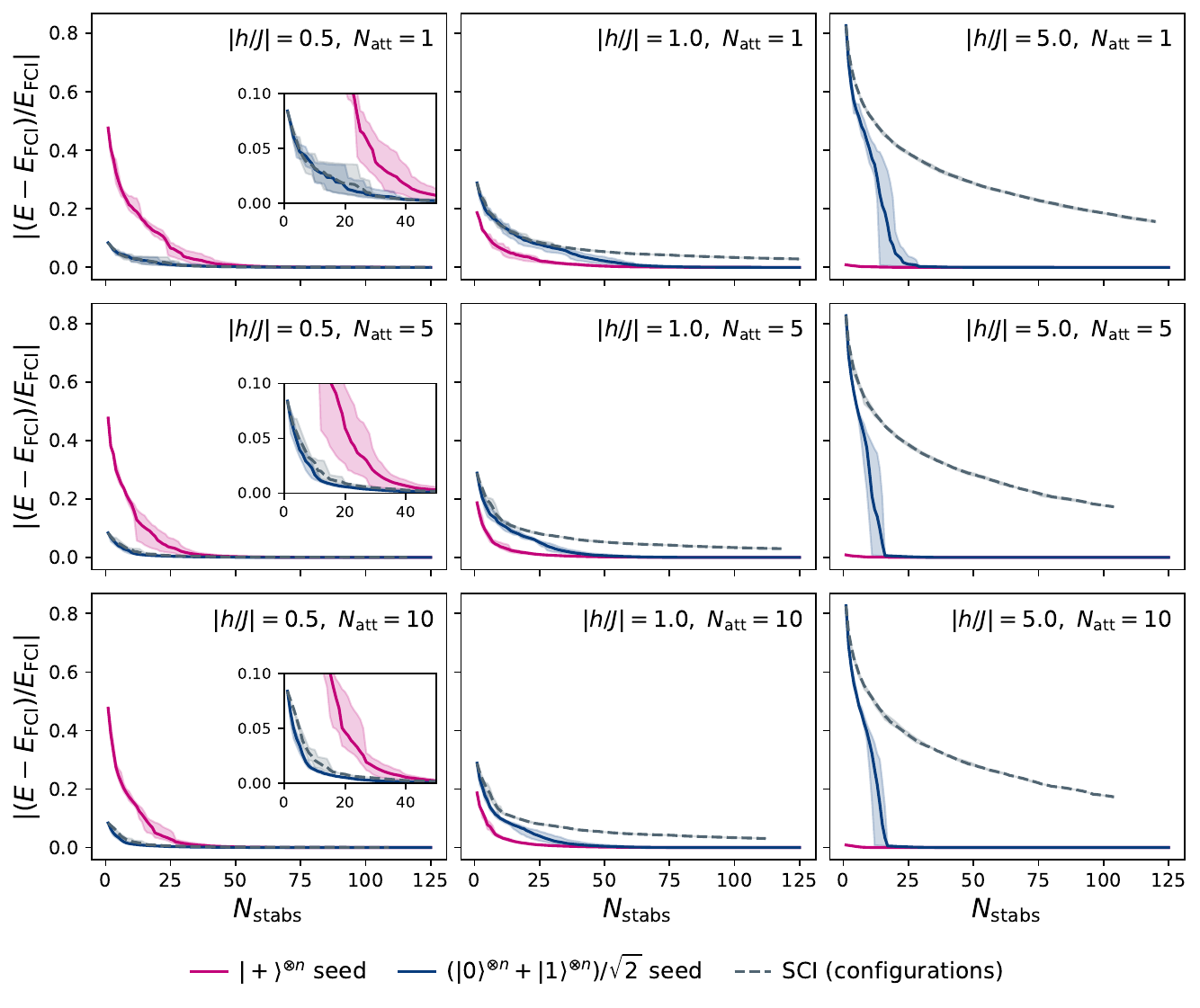}
    \caption{Convergence of relative energy error in the SSI procedure for the 1D TFI chain with $n=8$, for different transverse field strengths (left to right) and candidate pool sizes at each iteration (top to bottom, controlled by $N_\text{att}$). Solid lines correspond to average values across five independent simulations to show stochastic variability, and shaded areas span the minimum to maximum range, with the different colours denoting different stabilizer initializations as shown. The dashed line corresponds to an equivalent SCI simulation where all states selected in the expansion are single $Z$-basis configurations. As $h$ increases, the optimal starting point shifts from a linear combination of ferromagnetic states to the uniform superposition state. Insets: Zoomed-in convergence for $h=0.5J$ to ensure comparison between SSI and SCI is visible.}
    \label{fig:tfi_8q}
\end{figure*}

We consider first the integrable 1D TFI Hamiltonian given in \Cref{eq:tfi}. We will limit ourselves to the ferromagnetic Ising model, with exchange interaction, $J <0$, and transverse field, $h < 0$. 
When $|h/J| \rightarrow 0$, the ground state is the fully ferromagnetic solution, so the single configurations $|0\ket^{\otimes n}$, $|1\ket^{\otimes n}$ or any linear combination of them will span the degenerate ground state manifold. As $|h/J| \rightarrow \infty$, the ground state tends to the uniform paramagnetic state, $|+\ket^{\otimes n}$. At $|h/J|=1$ in the thermodynamic limit there is a quantum phase transition between the ferromagnetic and paramagnetic phases, and so we expect finite-size signatures of this criticality and long-range correlations to be strongest near \(\lvert h/J\rvert=1\), making this an especially demanding regime between the two stabilizer limits. The advantage of stabilizer states over configurational basis states is obvious in the high $h$ regime, where an exact representation of the ground state solution requires one stabilizer, but a superposition over $2^n$ basis states~\cite{Oliviero2022}. Therefore, both Hamiltonian parameter limits tend to a single stabilizer state, and exhibit clear exponential advantage in the high $h$ limit, and so this is expected to be a favorable starting system for consideration of the scheme.

We first explore the performance of SSI for different values of the $N_\text{att}$ parameter, which controls candidate pool size, to assess whether the exploration algorithms proposed are effective in searching the candidate stabilizer space. Results for $n=8$ are shown in \Cref{fig:tfi_8q}, with similar trends observed for $n = 3-12$.
We also compare the rate of convergence of the SSI expansion where the initial single `seed' stabilizer for the algorithm is either the GHZ $\frac{1}{\sqrt{2}}(|0\ket^{\otimes n} + |1\ket^{\otimes n})$ (exact in the $h=0$ limit) or $|+\ket^{\otimes n}$ (exact in the $h\rightarrow \infty$ limit), to see the effect of these very different starting points in converging the algorithm in different parameter regimes. We note that for convergence the initial stabilizer state must have a contribution from the same symmetry sector as the ground state wavefunction, precluding e.g. $\frac{1}{\sqrt{2}}(|0\ket^{\otimes n} - |1\ket^{\otimes n})$, which is not in the even sector with respect to the spin-flip symmetry of the Hamiltonian. We also compare to the SCI analog of the algorithm where every state in the expansion is a simple configuration, to faithfully compare the rate of convergence with respect to the length of the expansions under the same ranking and selection criteria.

\begin{figure}
    \centering
    \includegraphics[width=\linewidth]{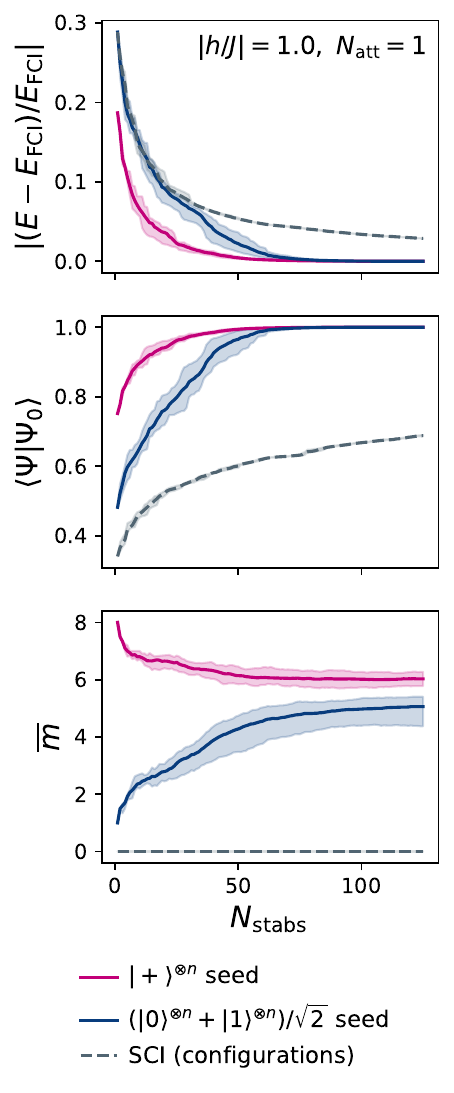}
    \caption{Convergence of SSI relative energy error, overlap magnitude with the ground state and average superposition degree, $m$, as the expansion increases for the 1D TFI model with $n=8$ and $|h/J| = 1$. Solid lines correspond to average values across five independent simulations and shaded areas span the minimum to maximum range.}
    \label{fig:tfi_8q_all}
\end{figure}

\begin{figure*}
    \centering
    \includegraphics[width=\linewidth]{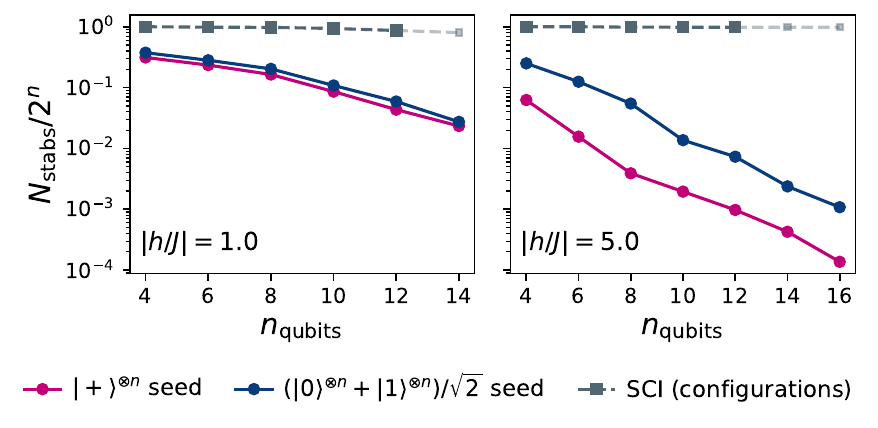}
    \caption{Normalized stabilizer expansion length needed to obtain convergence of the energy and fidelity with SSI for the 1D TFI model, with convergence defined as $\left|\frac{E - E_\mathrm{FCI}}{E_\mathrm{FCI}}\right| < 0.01$ and $1-|\bra\Psi_\text{SSI}|\Psi_\text{FCI}\ket| < 0.01$. The dashed lines corresponds to the number of states required within the configurational basis SCI procedure. The 14- and 16-qubit systems for this method are extrapolated due to computational cost. 
    }
    \label{fig:tfi_conv}
\end{figure*}

The results show a very weak dependence on $N_\text{att}$ and the size of the candidate pool at each step, implying an effective search of the stabilizer space to grow the expansion. Furthermore, relatively small variability within different runs is found. This allows us to use a low-cost $N_\text{att} = 1$ for the remainder of the results. At all field strengths, we find that the results converge to the ground state, and provided that an appropriate seed initialization is chosen, the stabilizer expansions improve on the rate of convergence of the configurational-based SCI results. The benefit in the rate of convergence over configurational expansions is greatest for larger field strengths as expected, with only small improvements as the field strength is reduced to the classical limit, where the ferromagnetic configurations dominate. However, significant compactification of the expansion is found even at the correlated $h=J$ regime near the phase transition compared to configurational results. We also find some dependence on the seed stabilizer for the algorithm. At $h=5J$ and $h=0.5J$, the use of the $|+\ket^{\otimes n}$ and $(|0\ket^{\otimes n} + |1\ket^{\otimes n})/\sqrt{2}$ seed stabilizers respectively do provide a more robust and slightly faster convergence, but seeding from the opposite stabilizer still does converge the algorithm to the correct state, albeit with a little more variability from the stochastic nature of the candidate selection algorithm, and a slightly slower rate. Nevertheless, all attempts converge to the correct results with monotonic variational energy improvements, and with far fewer than $2^n$ states. Even at $h=5J$, where the $(|0\ket^{\otimes n} + |1\ket^{\otimes n})/\sqrt{2}$ starts at the same energy as the SCI expansion, the SSI convergence is significantly faster than SCI.

We further analyze the resulting expansions in \Cref{fig:tfi_8q_all} at the critical $h=J$ parameter regime, where we also show the overlap magnitude with the exact ground state (fidelity) and average superposition degree of the stabilizer states selected. The convergence of the fidelity accentuates the disparity between the stabilizer and configurational expansions, clearly showing the improved convergence of the stabilizer representations to a faithful ground state. The average superposition degree of the stabilizers, $\bar{m}$, evolves over the course of the simulation from their different seeds towards an average of six, implying that the stabilizers selected are (on average) superpositions of $\approx 64$ configurations. This highlights that the stabilizer expansions are achieving their compression and selecting stabilizers with broad configurational basis support that are unavailable to a configuration-only expansion, validating the premise of the work.

We can also consider the system size dependence, with \Cref{fig:tfi_conv} showing the normalized stabilizer expansion length (as a fraction of the full Hilbert space size) required for convergence of the SSI algorithm to a desired tolerance (in this case both relative energy error and infidelity of the state below 0.01) in the $h=J$ and $h=5J$ parameter regimes. Encouragingly, we find decreasing fractions of the Hilbert space are needed to represent the wave function as the number of sites increase, with this trend clearer and steeper for the $h=5J$ system, which also allows calculations to converge rapidly for larger numbers of qubits in this regime. This contrasts with the configurational expansion, where very little of the configurational space can be neglected, even as the system size increases, with almost the entire space needing to be sampled for convergence to this accuracy. 

\subsection{Heisenberg XXZ model} \label{sec:XXZ}

\begin{figure}
    \centering
    \includegraphics[width=\linewidth]{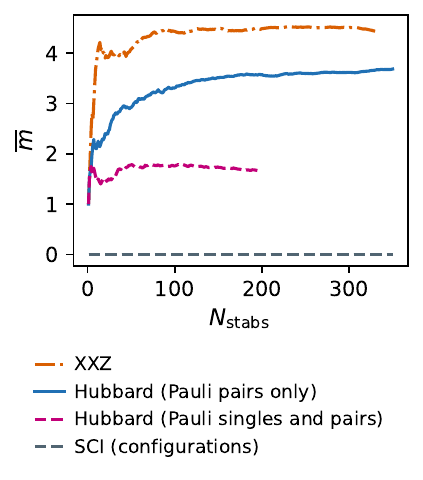}
    \caption{Evolution of the average stabilizer superposition degree during an SSI calculation for the 12-qubit XXZ and Hubbard models (see Sec.~\ref{sec:hubbard}), using symmetry-preserving candidate selection, while also including single Pauli-generated candidates in the Hubbard results. This average superposition degree is defined as $\bar{m} = M_t^{-1} \sum_\alpha m(\sigma_\alpha)$, with $m(\sigma_\alpha)$ defined in \Cref{eq:supdeg}.}
    \label{fig:xxz_superposition}
\end{figure}

The TFI model is a natural target for the SSI approach, as the ground state wave function at high field strengths has exponential support on the Hilbert state configurational basis states, but is close to a single stabilizer state. We next consider the Heisenberg XXZ Hamiltonian with open boundary conditions,
\begin{equation}
       \hat H_\text{XXZ} = J_{xy} \sum_{i =0}^{n-2} (X_iX_{i+1} + Y_iY_{i+1}) + J_z \sum_{i=0}^{n-2} Z_iZ_{i+1}. 
\end{equation}
Since $[\hat H_\text{XXZ}, \hat S_z] = 0$, with $\hat S_z =\frac{1}{2}\sum_{i=0}^{n-1}Z_i$ the total spin magnetization operator, the Hamiltonian can be block-diagonalized in sectors of fixed total magnetization quantum number, $S_z$. We consider the parameters $J_{xy}, J_z> 0$ and an even number of sites, for which the ground state has $S_z = 0$~\cite{SciPostPhys.7.2.023}. 

Although an $S_z=0$ state can be represented using stabilizers that individually contain components outside this sector, we impose the stronger symmetry-adapted restriction that every selected stabilizer has computational basis support entirely within the Hamming-weight-$n/2$ sector. Thus each selected state satisfies $\hat S_z|\sigma_\alpha\rangle=0$, and every resulting SSI expansion remains exactly within the target sector.
However, the candidate stabilizer proposal scheme described in Sec.~\ref{sec:ssi_algo} does not generally preserve either the magnetization or its variance, even when starting from a stabilizer with definite $S_z$.

\begin{figure*}
    \centering
    \includegraphics[width=\linewidth]{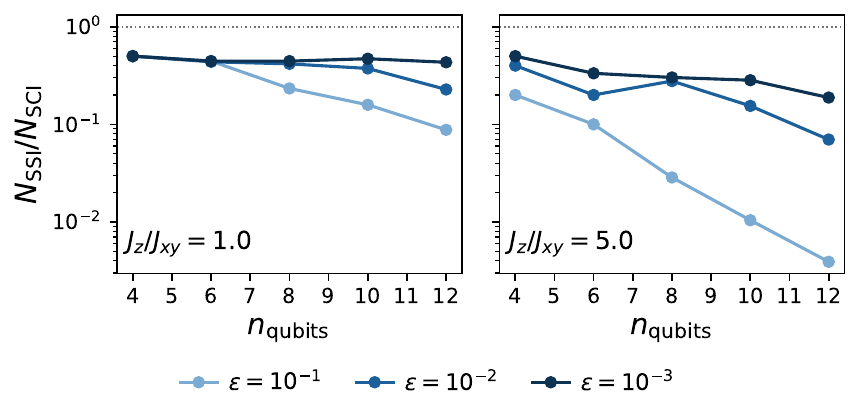}
    \caption{Ratio of stabilizer to configurational basis state expansion length required to achieve a fidelity threshold of $\epsilon$ to the exact ground state for the 1D XXZ model at different lattice sizes and values of $J_z/J_{xy}$. Convergence of the algorithm is defined as the expansion length for which $1-|\langle \Psi_\text{SSI}|\Psi_\text{FCI}\rangle| < \epsilon$.}
    \label{fig:xxz_conv}
\end{figure*}

Given the simple form of the XXZ Hamiltonian, we can adapt the candidate stabilizer selection scheme to preserve Hamming weight of the configurational support for all selected stabilizers, without leading to a constant superposition degree during the propagation. In order to do so, we include a set of sufficient conditions on the selected (signed) Pauli pairs $\hat P_1$ and $\hat P_2$ from the Hamiltonian to filter candidates to preserve $S_z$ symmetry, with the pair rejected and reselected if conditions are not met. These conditions are
\begin{enumerate}
    \item Both $\hat P_1$ and $\hat P_2$ are diagonal terms of the form
$Z_iZ_{i+1}$.
    \item The pair is $\{\hat P_1,\hat P_2\} = \{X_iX_{i+1},Y_iY_{i+1}\}$ on the same bond and with the same relative sign. In this case, locally aligned $\uparrow \uparrow$ or $\downarrow \downarrow$ configurations are annihilated, while $\uparrow \downarrow$ and $\downarrow \uparrow$ configurations in the superposition of the stabilizer are exchanged to preserve Hamming weight.
    \item Except for the same-bond case above, an off-diagonal Pauli selected as part of a pair, $\hat P \in\{X_iX_{i+1},Y_iY_{i+1}\}$, must satisfy $-Z_iZ_{i+1}\in\mathcal S(\sigma)$. This condition ensures that every configuration in the support of the parent stabilizer is locally anti-aligned on the relevant bond, so that the action of $\hat P$ exchanges $\uparrow \downarrow$ and $\downarrow \uparrow$ without changing the Hamming weight.
\end{enumerate}
As before, for accepted selections, we also check they do not annihilate the target stabilizer. If $\hat P_1$, $\hat P_2$ anticommute, we accept the candidate. If they commute, we only accept if and only if $-\hat P_1 \hat P_2 \notin \mathcal{S}(\sigma)$. Otherwise, we reject the selected pair and reselect.
We numerically find that applying these adaptations to explicitly enforce $S_z$ conservation for every candidate stabilizer results in a faster converging stabilizer expansion of the ground state of the model, compared to letting the symmetry naturally emerge from the expansion. This selection scheme still allows the superposition degree to change as the stabilizer expansion grows. An example for the 8-site XXZ model is shown in \Cref{fig:xxz_superposition}. The average superposition degree is however lower than the equivalent TFI model, which can be attributed to both the smaller Hilbert space size in the Heisenberg ground state sector, as well as the $S_z=0$ constraints on the computational support of each stabilizer.

We also require an appropriate initial stabilizer seed for the SSI which possesses the desired $S_z$, and choose the antiferromagnetic linear combination of the two spin-flipped N{\'e}el states, $|N\ket = |01\ket^{\otimes n/2}$ and $|\bar N\ket = |10\ket^{\otimes n/2}$ to preserve global spin-flip symmetry. We note that for $n = 4k$, the ground state symmetry sector corresponds to $|\sigma_1\ket = (|N\ket + |\overline{N}\ket)/\sqrt{2}$, while for $n=4k+2$, it corresponds to $|\sigma_1\ket = (|N\ket - |\overline{N}\ket)/\sqrt{2}$, according to the Marshall sign rule structure, and so choose the relative phase of these two states in the seed appropriately.

The convergence of the stabilizer expansion size for this model to achieve a given accuracy in the fidelity of the ground state, $\epsilon$, relative to the rank of a pure configurational state expansion, is shown in \Cref{fig:xxz_conv}. Across the sizes and tolerances tested, SSI reaches the overlap criterion with fewer retained states than the configurational expansion. 
The ratio of stabilizers to configurations we find to approximate the XXZ ground state to a fixed fidelity decreases as the system size increases, showing an increasing compression of the ground state correlations in the stabilizer representations and implying that they can increasingly benefit from the more extensive superposition structures available in the stabilizer expansions. This improvement with system size is clearest for larger values of $J_z/J_{xy}$, closer to the Ising limit with better alignment with the stabilizer set. As expected, as the target fidelity threshold is tightened to approach the exact ground state there is a reduction in the stabilizer compression, as both expansions tend towards a requirement to span the complete Hilbert space. The stabilizer compression we find in these limits tends towards one half regardless of system size, consistent with pairing configurations related by global spin flip being described within the stabilizer space.
 
The convergence of the energy with system size is broadly similar, but the presence of many low-lying excited states in the model makes the differences harder to discern, with flatter convergence profiles.

\subsection{Hubbard model} \label{sec:hubbard}

We finally consider the one-dimensional spinful Hubbard model with open boundary conditions in the site basis, given by
\begin{equation}
\begin{split}
    \hat H_{\mathrm{hub}} = &-t\sum_{i=0}^{L-2}\sum_{\sigma=\uparrow,\downarrow}
\left(\hat c_{i\sigma}^{\dagger}\hat c_{i+1,\sigma}
+\mathrm{H.c.}\right)\\
&+ U\sum_{i=0}^{L-1} \hat n_{i\uparrow}\hat n_{i\downarrow},
\end{split}
\end{equation}
where $t$ is the nearest-neighbour hopping amplitude, $U$ is the on-site interaction strength, and $\hat n_{i\sigma}=\hat c_{i\sigma}^{\dagger}\hat c_{i\sigma}$. For the repulsive model considered here, $U/t$ controls the interaction strength relative to hopping. We will focus on even numbers of sites, and half-filling with $S_z =0$. Even the simplest two-site Hubbard model has been recently shown to display meaningful non-stabilizerness (defined as the minimal one-norm of an exact stabilizer expansion) away from the $U/t \rightarrow 0$ and $U/t \rightarrow \infty$ limits~\cite{Zavatti2026}. Beyond two sites, neither $U/t$ limit can necessarily be represented by a simple Pauli stabilizer state, although for $U=0$ there exists a simple Fermionic Gaussian unitary state solution. Therefore, we expect this system to prove challenging for SSI. 

 We first map this Fermionic model to Pauli operators via a Jordan-Wigner transformation, resulting in one qubit per spin-orbital ($n=2L$), ordering the qubits by spin channel and then site label. Other fermion-to-qubit mappings, qubit tapering, as well as a Majorana stabilizer representation to conserve fermionic parity symmetry can be considered in the future~\cite{bravyi2017taperingqubitssimulatefermionic,Bravyi_2010}. Similar to the XXZ system, additional symmetries are present in the Hamiltonian which are not natively preserved in the stabilizer definition, most importantly electron number and $S_z$, so we again develop a filtering approach to rigorously guarantee that no broken $N_\text{elec}$- or $S_z$-symmetry configurations appear in the support of the proposed stabilizer states, with an analogous scheme to the one described in Sec.~\ref{sec:XXZ}. Beyond these symmetries, other discrete symmetries exist in the Hamiltonian, including the mirror-plane symmetry, global spin-flip and particle-hole symmetry. These should result in at least a factor of two compression in the stabilizer expansion length compared to the configurational basis, since pairs of equal-weight configurations preserving these symmetries can be combined similar to the analysis of Sec.~\ref{sec:motivation}. We initialize the SSI algorithm for this model with a symmetry-adapted two-configuration N{\'e}el state stabilizer. We also considered alternative seed stabilizers such as valence-bond or singlet-paired states, but results from these did not prove significantly changed.

 Unlike the spin-model benchmarks, our initial investigations on this model exhibited poor compression of the SSI method compared to the configurational basis. In particular, for the larger systems and higher desired accuracies, the SSI expansion length was even found to exceed the equivalent SCI approach (see \Cref{fig:app_hub}). Because computational basis configurations are themselves stabilizer states, these ratios clearly reflect limitations of the present selection procedure rather than a fundamental lower bound on stabilizer representations.

\begin{table}[]
    \centering
    \begin{tabular}{|c|c|c|c|}
         \hline
         $n_\text{qubit}$ & $n_\text{Hilbert}$ & $n_\text{stabs}$ & $\bar m$\\
         \hline
         4& 4 & 2 & 1.00\\
         8& 36 & 10  & 1.60\\
         12 & 400 & 104 & 1.92\\
         \hline
    \end{tabular}
    \caption{Upper bound on the exact stabilizer rank of the ground state of Hubbard chains, obtained via a greedy selection of orthogonal stabilizers. The found stabilizer rank is given by $n_\mathrm{stabs}$, while $n_\textrm{Hilbert}$ denotes the dimensionality of the $N_\mathrm{elec}$ and $S_z=0$ symmetry allowed sector of the configurational Hilbert space. $\bar{m}$ denotes the average superposition degree of the found expansion. Results apply to all $U>0$.}
    \label{tab:hub_fci_rank}
\end{table}

\begin{figure*}
    \centering
    \includegraphics[width=\linewidth]{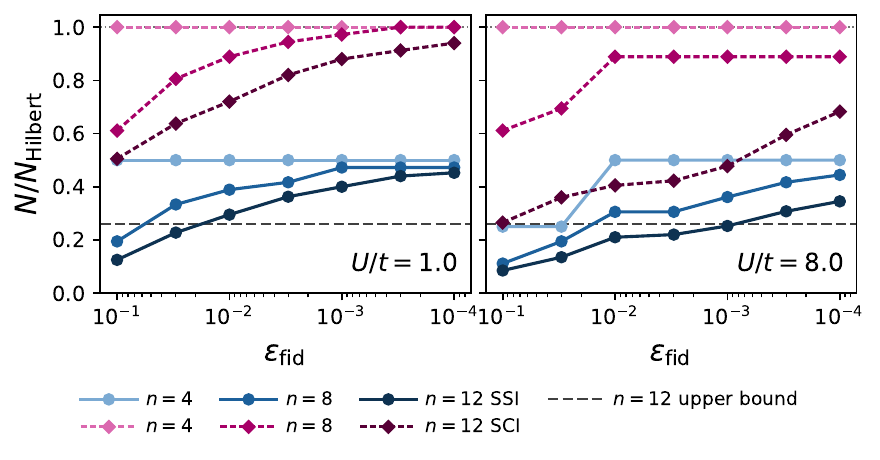}
        \caption{Length of stabilizer (solid lines) or configurational (dashed lines) basis expansions compared to the full Hilbert space, required to achieve a given infidelity from the SSI / SCI algorithm, $\epsilon_\text{fid}$, at different values of $U/t$, and qubit number $n$, for the 1D Hubbard chain. }
    \label{fig:hub_conv}
\end{figure*}

To analyze this further, and distinguish a fundamental inadequacy in the stabilizer representation from an algorithmic limitation in the SSI selection, we seek a procedure to bound the exact stabilizer rank of the ground state wave function for these systems. While determining the exact stabilizer rank of a given state is a challenging and open problem~\cite{Bravyi2019, Peleg2022,Mehraban2024}, we design a simple greedy algorithm to at least provide a loose upper bound to this rank via an orthogonal stabilizer expansion. This proceeds by simply grouping configurations from the exact ground state with the same amplitude, and reducing them to a single stabilizer if they form the required affine subset of $\mathbb F_2^n$ in their bit string representations. Grouped configurations are then removed from the residual, and the process repeated until no more stabilizer groupings can be found. The resulting stabilizer expansion is manifestly orthogonal, since the underlying configurational support for each is disjoint. This results in a crude upper bound for the stabilizer rank, with more optimal stabilizer decompositions likely to exploit non-orthogonality of these expansions. The bound on these stabilizer ranks, $n_\textrm{stabs}$, and the average superposition degree of the found expansion, $\bar{m}$, is given in \Cref{tab:hub_fci_rank}. This shows that to represent the {\em exact} ground state of the 12 qubit Hubbard model, an expansion of at most 104 stabilizers is needed (approximately a quarter of the Hilbert space size), with each stabilizer therefore on average having support on approximately four configurations.

We found that even these loose upper bounds on the stabilizer rank for the exact state were quite significantly below the expansion lengths found by our initial SSI procedure for approximate ground states. For instance, in the $n_\textrm{qubit}=8$ model, we found 20 stabilizers required for $U/t=1$ and 27 for $U/t=8$, despite the the upper bound on the stabilizer rank being only 10. Delving deeper into this discrepancy, one notable feature of the expansions constructed from the exact state is that the average superposition degree, $\bar{m}$, is significantly lower than the one found by the SSI procedure, shown in Fig.~\ref{fig:xxz_superposition}. This pointed to a potential failure of the current SSI approach, since correcting for the inclusion of stabilizers which group together configurations with very different final amplitudes requires many more stabilizers to mix to recover the desired result. 

Within the proposed SSI selection algorithm, the choice to combine the action of two Paulis to generate new stabilizers biases these candidates towards different superposition degrees from their parent stabilizers. This leads to a rapid search (and inclusion) of states with very different superposition degree. Therefore, informed by the analysis above, we mitigate the growth of $m$ in the expansion by also including stabilizer candidates generated by single Pauli strings selected by the Hamiltonian to the pool. For a parent stabilizer $|\sigma\ket$ and selected Paulis $\hat P_1$ and $\hat P_2$, we include $\left\{\hat P_1|\sigma\ket, \hat P_2|\sigma\ket, \frac{1}{\sqrt{2}}\left.\left.\left(s_1\hat P_1 + s_2\hat P_2\right)\right|\sigma\right\rangle\right\}$ into the candidate pool. This leads to the development of SSI expansions in the Hubbard model with lower $\bar{m}$ values, and improved compactness of the stabilizer expansions found via SSI, as shown in \Cref{fig:xxz_superposition}. Comparing this modified selection procedure to the SCI expansion lengths in \Cref{fig:hub_conv} demonstrates that we at least bound the advantage in expansion length over SCI by at least the anticipated factor of two, while now more clearly improving on this if lower fidelity is required. At higher desired fidelities in the SSI expansion however, we still converge to expansion lengths above the upper bound reported in \Cref{tab:hub_fci_rank} obtained by starting from the exact state. This clearly points to the sensitivity of the precise candidate selection method, rather than a fundamental limitation of stabilizer expansions as key to further improvements. Being able to effectively search for approximate stabilizer expansions for a target fermionic state where the exact state is unknown is a continuing research direction.

\section{Conclusion} \label{sec:conclusion}

We have introduced stabilizer selected interaction (SSI), an SCI-inspired variational method that iteratively diagonalizes a subspace Hamiltonian from a basis of selected stabilizer states. The approach systematically enlarges the selected space via an efficient classical heuristic for searching over candidate stabilizers. Because an individual stabilizer can compactly encode a coherent superposition of many configurations while retaining classically tractable overlap and Hamiltonian matrix-element evaluation, SSI aims to provide a more compact alternative to configuration selected expansions. The key question of this work was to validate whether a classical heuristic could efficiently search the stabilizer space for these compact expansions in a black-box fashion, and achieve improvements on configurational based equivalent heuristics, comparing rigorously against exact diagonalization results. 

The results demonstrate substantial but model-dependent promise. For quantum spin-systems like the TFI model, we have shown that this approach can indeed achieve substantially more compact expansions, converging to the exact ground state, and with a performance advantage which clearly grows with system size. Positive results were also achieved with the XXZ Heisenberg model, where adaptations of the search heuristic ensured that $S_z$ symmetry was conserved throughout the evolution, enabling SSI to successfully recover the ground state wavefunction at an accelerated rate compared to a configurational expansion. This was extended to the fermionic Hubbard model, which proved more challenging, and yielded little compression in the expansion by changing from a configurational to stabilizer underlying representation. This outcome exposed a limitation of the search heuristic rather than establishing the absence of compact stabilizer representations for fermionic ground states. 

Analysis against an upper bound on the exact stabilizer rank of the target state pointed to this being due to the constrained selection scheme precluding an effective search of the stabilizer space or allowing the superposition properties of the stabilizers to be exploited. A simple change to expand the candidate pool at each step yielded improvements, though we know we are still far from the optimal algorithm to search the stabilizer space in the absence of a provided and known target state. 

While this therefore represents a first proof-of-principle of a general and efficient classical scheme for iteratively converging variational stabilizer expansions in many-body physics, there is clearly further work required to expand it to fermions and, hopefully, \textit{ab-initio} Hamiltonians. In particular, this work deliberately focuses on the compactness of the resulting expansion, rather than other relevant metrics such as the overall time-to-solution to determine any overall algorithmic advantage. These will have to be explored in the future, and will involve further algorithmic design to maximize the efficiency of the implementation. At this point, we can scale up the analysis of system sizes beyond exact results, to clarify the empirically observed scaling advantages, delineating the scope of polynomial from exponential advantage and whether the compression persists to larger systems. 

There are also clearly algorithmic improvements which are required. The first requirement is to develop more flexible selection routines which can exploit symmetry without biasing the superposition degree of candidate stabilizers, for which penalty or symmetry projection operators could be envisaged. Secondly, one must consider a robust selection mechanism for the initial state, where relying on alternative classical heuristics could be found. The quality of the stabilizer expansion also depends on the structure of the wavefunction in the Hilbert space, which is highly basis set dependent. Combining the stabilizer expansion with optimized orbitals or fermionic Gaussian unitaries to optimize the compression of the resulting stabilizer expansion will be an interesting further step. Finally, while the method presented here is fully classical, the use of stabilizer states as a more effective basis for hybrid quantum-classical SQD/QSCI methods will also be explored as the subject of further work.

\section*{Acknowledgements}
MAF acknowledges support from Peterhouse, Cambridge through a Research Fellowship and from Downing College, Cambridge through the Kim and Julianna Silverman Research Fellowship. GHB acknowledges support from EPSRC via award UKRI6209. The authors thank Max Marvell and Nick Mayhall for insightful discussions.

\section*{Author Contributions}
The authors conceptualised and developed the project together. MAF developed the required software, generated and analysed data. Both authors contributed equally to interpretation of results and writing the manuscript. An LLM was used for code testing and optimisation, as well as figure generation.

\bibliography{./main}
\newpage
\appendix

\onecolumngrid

\section{Further stabilizer state simulation algorithms}
\label{app:code}
In this appendix we give pseudocode algorithms for obtaining phased overlap of stabilizer states (Algorithm \ref{alg:overlap}), the phased conjugation by a general Clifford circuit (Algorithm \ref{alg:conjcirc}), as well as the phase of a Hadamard gate conjugation of a stabilizer generator matrix (Algorithm \ref{alg:hadamard}). 

\begin{algorithm}[!h]
    \caption{Exact complex overlap between stabilizer states}
    \label{alg:overlap} \footnotesize
    \begin{algorithmic}[1]
        \Require Canonical stabilizer generator matrices $\mathcal{M}^\phi$ and $\mathcal{M}^\psi$, with rows $P_i$ and $R_i$ respectively.
        \Ensure Overlap $\bra \phi | \psi \ket$.
        \Statex \hspace{-5mm} $\Rightarrow$ {\tt{STABCIRC}}($\mathcal{M}^\phi$) returns the Clifford circuit such that $\mathcal{C}|\phi\ket = |b\ket$. and $\mathcal{M}^{b}$, the canonical generator matrix for $b$. 
        \Statex \hspace{-5mm} $\Rightarrow$
        {\tt{CONJCIRC}}($\mathcal{M}^\phi, \mathcal{C},\mathcal{M}^{b})$ returns $\mathcal{M}^{\phi'}$ for $|\phi'\ket =\mathcal{C}|\phi\ket$, a boolean $f$ storing whether $\bra{b|\phi'\ket} \neq 0$ and the complex phase $p$ associated with $b$ in $\phi'$. Details in \Cref{alg:conjcirc}.
        \Statex \hspace{-5mm} $\Rightarrow$ {\tt{INNER\_PROD}} ($\mathcal{M}^{\phi'}, \mathcal{M}^{\psi'}$) returns {\tt{is\_zero}} and $k$, such that $|\bra\psi|\phi\ket| = 0 \text{ or }2^{-k/2}$. See \cite{Garcia2014} for details.
        \Statex
        \State $\mathcal{C}, \mathcal{M}^{|b\ket} \leftarrow$ {\tt{STABCIRC}}($\mathcal{M}^\phi$)
        \State $\mathcal{M}^{|\psi'\ket},p_\psi, f_1 \leftarrow$ {\tt{CONJCIRC}}($\mathcal{M}^\psi, \mathcal{C},\mathcal{M}^{|b\ket}$)
        \State $\mathcal{M}^{|\phi'\ket},p_\phi, f_2 \leftarrow$ {\tt{CONJCIRC}}($\mathcal{M}^\phi, \mathcal{C},\mathcal{M}^{|b\ket}$)
        \Comment{$|\phi'\ket = |b\ket$}
        \State \tt{is\_zero}, $k \leftarrow$ {\tt{INNER\_PROD}} ($\mathcal{M}^{|\phi'\ket}, \mathcal{M}^{|\psi'\ket}$)
        \State \tt{if is\_zero}
        \State \hspace{5mm} \tt{return} 0
        \State \tt{else if } $f_1$ \tt{and} $f_2$
        \Comment{Either is\_zero or ($f_1$ and $f_2$) must hold.}
        \State \hspace{5mm} \tt{return} $p_\phi^*\times p_\psi\times2^{-k/2}$
    \end{algorithmic}
\end{algorithm}
\begin{algorithm}[!t]
    \caption{Phased conjugation of a stabilizer state by a Clifford circuit ({\tt CONJCIRC})}
    \label{alg:conjcirc} \footnotesize
    \begin{algorithmic}[1]
        \Require Generator matrix $\mathcal{M}^\psi$; Clifford circuit $\mathcal{C}$; generator matrix $\mathcal{M}^b$ of the target basis state $|b\ket \propto \mathcal{C}|\phi\ket$.
        \Ensure $\mathcal{M}^{\psi'}$ for $|\psi'\ket = \mathcal{C}|\psi\ket$; the phase $p$ of $|b\ket$ in $|\psi'\ket$; a flag $f$, true if $\bra b|\psi'\ket \neq 0$.
        \Statex \hspace{-5mm} $\Rightarrow$ {\tt FIND\_BASIS\_STATE}($\mathcal{M}$) returns the canonical basis state contained in the state with generator matrix $\mathcal{M}$.
        \Statex \hspace{-5mm} $\Rightarrow$ {\tt HADAMARD\_PHASE}($\mathcal{M}, \mathcal{M}^t, j$) returns the phase factor and the updated tracked basis state after applying $H_j$. Details in \Cref{alg:hadamard}.
        \Statex \hspace{-5mm} $\Rightarrow$ {\tt ConjugateG}($\mathcal{M}$) returns the result of conjugating the generator matrix $\mathcal{M}$ by Clifford operation $G$.
        \Statex \hspace{-5mm} $\Rightarrow$ {\tt RELPHASE}($\mathcal{M}^\phi, \mathcal{M}^{b}$) returns {\tt found}, true if $|b\ket$ is in the state, and $lp = \bra b|\phi\ket / \bra c|\phi\ket$, the phase of $|b\ket$ relative to $|c\ket$, the canonical basis state.
        \Statex
        \State $\mathcal{M} \gets \mathcal{M}^\psi$,\ \ $p \gets 1$,\ \ $\mathcal{M}^t \gets$ {\tt FIND\_BASIS\_STATE}($\mathcal{M}^\psi$)
            \Comment{Track the initial canonical basis state $|t\ket$ and its phase $p$}
        \State $s_k \equiv$ sign of $Z_k$ in $\mathcal{M}^t$
            \Comment{$s_k = -1 \Leftrightarrow$ qubit $k$ is $|1\ket$ in $|t\ket$}
        \ForAll{gates $g \in \mathcal{C}$}
            \If{$g = H$ on qubit $j$}
                \State $p_H,\ \mathcal{M}^t \gets$ {\tt HADAMARD\_PHASE}($\mathcal{M}, \mathcal{M}^t, j$);\ \ $p \gets p\cdot p_H$
                \State $\mathcal{M} \gets$ {\tt ConjugateH}($\mathcal{M}, j$)
            \ElsIf{$g = P$ on qubit $j$}
                \State $\mathcal{M} \gets$ {\tt ConjugateP}($\mathcal{M}, j$);\ \ \textbf{if} $s_j = -1$ \textbf{then} $p \gets i\,p$
            \ElsIf{$g = \mathrm{CNOT}$ with control $j$, target $k$}
                \State $\mathcal{M} \gets$ {\tt ConjugateCNOT}($\mathcal{M}, j, k$);\ \ \textbf{if} $s_j = -1$ \textbf{then} $s_k \gets -s_k$
            \ElsIf{$g = \mathrm{CZ}$ on qubits $j, k$}
                \State $\mathcal{M} \gets$ {\tt ConjugateCZ}($\mathcal{M}, j, k$);\ \ \textbf{if} $s_j = s_k = -1$ \textbf{then} $p \gets -p$
            \ElsIf{$g = X$ on qubit $j$}
                \State $\mathcal{M} \gets$ {\tt ConjugateX}($\mathcal{M}, j$);\ \ $s_j \gets -s_j$
            \EndIf
        \EndFor
\Comment{$p$ is currently the phase of the tracked state $|t\ket$, not the target $|b\ket$.}
        \State {\tt found}$_1$, $lp_1 \gets$ {\tt RELPHASE}($\mathcal{M}, \mathcal{M}^t$)
            \Comment{$lp_1 = \bra t|\psi'\ket / \bra c|\psi'\ket$}
        \State {\tt found}$_2$, $lp_2 \gets$ {\tt RELPHASE}($\mathcal{M}, \mathcal{M}^b$)
            \Comment{$lp_2 = \bra b|\psi'\ket / \bra c|\psi'\ket$}
        \State $p \gets p \cdot lp_2 / lp_1$
            \Comment{Now $p$ is the phase of $|b\ket$ in $|\psi'\ket$}
        \State \Return $\mathcal{M},\ p,\ (${\tt found}$_1$ \textbf{and} {\tt found}$_2)$
            \Comment{{\tt found}$_1$ is always true; {\tt found}$_2$ false $\Leftrightarrow \bra b|\psi'\ket = 0$}
    \end{algorithmic}
\end{algorithm}

\begin{algorithm}
\caption{Phase update for a Hadamard gate ({\tt HADAMARD\_PHASE})}
\label{alg:hadamard}
\begin{algorithmic}[1]
\Require Generator matrix $\mathcal{M}$; basis state generator matrix $\mathcal{M}^b$; qubit $j$;
\Ensure Phase factor $p_H$ and updated $\mathcal{M}^b$ such that the new $|b\ket$ is in $H_j|\psi\ket$ with additional phase $p_H$.
\Statex
    \State $s_j \equiv$ sign of $Z_j$ in $\mathcal{M}^b$
        \Comment{$s_j = -1 \Leftrightarrow$ qubit $j$ is $|1\ket$ in $|b\ket$}
    \State $\mathcal{M}^{b'} \gets \mathcal{M}^b$
    \State $s_j' \gets -s_j'$
    \State ${\tt found1}, \ell_1 \gets {\tt{RELPHASE}}(\mathcal{M}, \mathcal{M}^b)$
    \State ${\tt found2}, \ell_2 \gets {\tt{RELPHASE}}(\mathcal{M}, \mathcal{M}^{b'})$
    \If{{\bf not} ({\tt found1} {\bf and} {\tt found2})}
        \State $\ell \gets 0$
    \Else
    \State $\ell \gets \ell_2 / \ell_1$
    \EndIf
     \Statex \Comment{$H_j(|x\ket + \ell|\bar x\ket) \propto (s_j + \ell)|x\ket + (1 - s_j\ell)|\bar x\ket$, where $|\bar x\ket = X_j|x\ket$}
    \If{$l = -s_j$} \Comment{$|x\ket$ is annihilated, so we track $|\bar x\ket$}
        \State $s_j \gets -s_j$
        \State $p_H \gets 1$
    \Else
        \State $p_H \gets(s_j + \ell)/|s_j + \ell|$ 
    \EndIf
\State \Return $p_H, \mathcal{M}^b$
\end{algorithmic}
\end{algorithm}

\FloatBarrier
\setcounter{figure}{0}
\renewcommand{\thefigure}{B\arabic{figure}}
\section{Supplementary Figures}
\label{app:sup_fig}

\begin{figure}[!h]
    \centering
    \includegraphics[width=\linewidth]{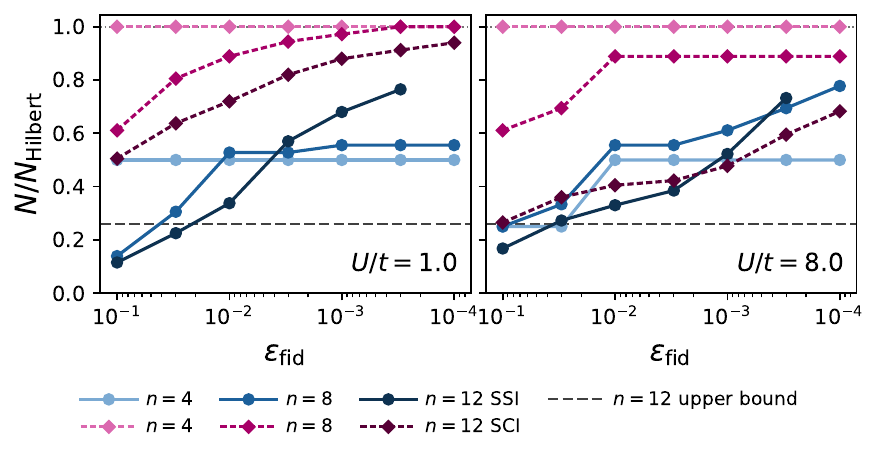}
        \caption{Length of stabilizer (solid lines) or configurational (dashed lines) basis expansions compared to the full Hilbert space, required to achieve a given infidelity from the SSI / SCI algorithm, $\epsilon_\text{fid}$, at different values of $U/t$, and qubit number $n$, for the 1D Hubbard chain. In contrast to Fig.~\ref{fig:hub_conv}, this initially tested SSI candidate proposal scheme only considered the action of pairs of Pauli operators to select new stabilizer states, with its poorer performance motivating the improved scheme of the main text which resulted in more compact stabilizer expansions. }
    \label{fig:app_hub}
\end{figure}

\Cref{fig:app_hub} shows the expansion length relative to the full Hilbert space in the Hubbard model for the initial SSI implementation, prior to the expansion of the candidate pool as described in the main text, as well as the comparable SCI method. For large systems, the SSI expansion exceeds the length of the SCI expansion needed for the same infidelity, motivating the improvements to the candidate selection procedure, leading to the improved results as shown in Fig.~\ref{fig:hub_conv}.

\end{document}